\documentclass[aps,pra,twocolumn,superscriptaddress,amsmath,amssymb,floatfix]{revtex4-2}

\usepackage{bm}
\usepackage{graphicx}
\usepackage{mathtools}
\usepackage{hyperref}
\hypersetup{colorlinks=true,linkcolor=blue,citecolor=blue,urlcolor=blue}

\newcommand{\dg}{^\dagger}
\newcommand{\Ima}{\operatorname{Im}}
\newcommand{\ad}{a\dg}
\newcommand{\ket}[1]{|#1\rangle}
\newcommand{\bra}[1]{\langle#1|}
\newcommand{\Ai}{\operatorname{Ai}}
\newcommand{\ketbra}[2]{\left|#1\right\rangle\!\left\langle#2\right|}
\providecommand{\braket}[2]{\langle#1|#2\rangle}
\newcommand{\melem}[3]{\langle#1|#2|#3\rangle}
\newcommand{\Rea}{\operatorname{Re}}

\begin{document}

\title{Exact Fock Amplitudes of the Cubic-Phase Gate on Arbitrary Pure Gaussian States}

\author{Ghasem Asadi Cordshooli}
\email{ghascor@iau.ac.ir}
\affiliation{Department of Physics, YI.C., Islamic Azad University, Tehran, Iran}

\date{\today}

\begin{abstract}
The cubic phase gate is the only non-Gaussian element of universal
continuous-variable quantum computation. Its Fock amplitudes are known in
closed form only when the input is a Fock state. For a Gaussian input the
usual route is to build the cubic quadrature in a truncated Fock space and
exponentiate it, which alters the generator before the exponential is taken.
A closed form is obtained here for every pure single-mode Gaussian input,
with the squeezing along any quadrature axis and with arbitrary displacement
and boost. It is a finite combination of the Airy function and its derivative,
and the argument is shared by all Fock indices, so a whole Fock profile costs
two Airy evaluations. A recurrence is also derived that builds each
coefficient from the four preceding ones, together with the regime in which it
is stable.
Evaluating the closed form at the weak nonlinearities of current experiments
requires care, since its terms grow large and alternate in sign. The precision
this demands is quantified, and the truncated construction is benchmarked
against the exact result, whose error at cutoffs in common use is found to be
of the same size as the quantity being computed.
\end{abstract}
\maketitle

\section{Introduction}
\label{sec:intro}

Continuous-variable quantum computation (CVQC) encodes information in the
infinite-dimensional Hilbert space of harmonic oscillators, with the canonical
quadratures $\hat x$ and $\hat p$ as the continuous degrees of
freedom~\cite{Braunstein2005,Lloyd1999}. CVQC offers a route to scalable and
fault-tolerant processing across optical~\cite{Menicucci2006,Alexander2016,
Asavanant2021,Larsen2021} and mechanical~\cite{Zinatullin2021} platforms, and,
unlike discrete-variable architectures~\cite{Raussendorf2001,Ohliger2010}, it
permits deterministic generation of large-scale entanglement. Gaussian
operations alone are not enough for universal computation. Displacements,
squeezers, beam splitters and phase shifters preserve Gaussian states and are
fixed by the first two moments of the quadratures, so they can be simulated
efficiently on a classical computer~\cite{Bartlett2002}; they are also
insufficient for fault-tolerant error correction~\cite{Niset2009}. Universality
therefore requires a non-Gaussian element, and in one standard formulation of
optical continuous-variable computation the entire gate set consists of five
members. These are the Fourier transform, the displacement, the quadratic phase
gate, a two-mode entangling gate, and one cubic
gate~\cite{FurusawaVanLoock2011}. The first four are Gaussian and have been
realized unconditionally in the laboratory for more than a
decade~\cite{Ukai2011}. The fifth is the \emph{cubic phase gate}
\begin{equation}
V(\lambda)=e^{i\lambda \hat x^{3}},
\label{eq:gate}
\end{equation}
introduced by Gottesman, Kitaev and Preskill together with the oscillator
encoding of a qubit~\cite{GKP2001}. Its generator is cubic in the quadratures,
so it produces Wigner negativity and implements non-Clifford operations on
encoded qubits~\cite{Fukui2023,Baragiola2019}. No linear optical transformation
can supply it, since it demands a third-order nonlinearity acting on states of
very low optical power~\cite{GhoseSanders2007,Marek2011,Marshall2015}.

What the schemes consume is not the gate but the state it produces from a
squeezed vacuum, the \emph{finite-energy cubic phase state}
\begin{equation}
\ket{\psi}=V(\lambda)\,S(r)\ket{0},
\label{eq:cps0}
\end{equation}
which serves as the ancilla in gate-teleportation implementations of the
gate~\cite{GhoseSanders2007,Marshall2015,Zinatullin2021}, as a resource for
non-Gaussian entanglement~\cite{McConnell2024}, and in the preparation of cat
and GKP states~\cite{Eaton2019}. More generally the input carries a
displacement and a boost, and its squeezing need not lie along $\hat x$ or
$\hat p$, so that the object of interest is $c_n=\bra{n}V(\lambda)\ket{G}$ for
an arbitrary pure single-mode Gaussian $\ket{G}$. The Fock amplitudes are the
natural target here rather than the wave function or the Wigner function,
because photon-number statistics are what heralding and detection act on and
what the resource-theoretic measures of non-Gaussianity are computed
from~\cite{Albarelli2018,Genoni2010,Takagi2018}. This general family is the one
the applications call for, since the classical fidelity thresholds against
which the teleportation step is judged are themselves defined for arbitrary
pure single-mode Gaussian inputs~\cite{Chiribella2014}.

The gate has been demonstrated optically by nonlinear
feedforward~\cite{Sakaguchi2023}, the culmination of a line of proposals and
partial realizations~\cite{Menzies2009,Marek2011,Yukawa2013,Miyata2016,
Marek2018}, at a strength corresponding to $\lambda\approx0.17$ in the
convention of \eqref{eq:gate}, and the decomposition schemes built on that
demonstration use individual strengths at or below it, in places as small as
$\lambda\simeq0.035$~\cite{Budinger2024}. Two consequences bear on everything
below. The first is that the physically relevant regime is weak nonlinearity,
$\lambda\lesssim0.2$, which is where the evaluation of a closed form is most
delicate and where the precision analysis of Sec.~\ref{sec:stab} therefore
lives. The second is that the relevant object is a circuit rather than a single
gate, since those schemes chain tens to hundreds of them, so an amplitude error
that is tolerable in one gate may accumulate across the circuit.

In phase space the structure of the gate is understood. The Wigner function of
the ideal cubic phase state was identified as an Airy function by Ghose and
Sanders~\cite{GhoseSanders2007}, and Moore and Filip have shown that the action
of nonlinear phase gates on an arbitrary state is an \emph{Airy transform} of
the Wigner function~\cite{Moore2025}. The reason is elementary. The gate
contributes a cubic polynomial to the exponent of an integral, and the Airy
function is what a cubic exponent integrates to.

In the Fock basis the corresponding statement is known for the gate alone.
Miatto and Quesada reduce the matrix elements $\bra{m}V(\lambda)\ket{n}$ to the
Airy function and its derivative at the single argument $(3\lambda)^{-4/3}$, by
a recurrence and by a closed analytic form, with the higher derivatives
eliminated through $\Ai''(z)=z\Ai(z)$~\cite{MiattoQuesada2020}. Two things lie
outside that result. It is a Fock-to-Fock statement, and composing it with the
Fock expansion of a Gaussian does not deliver $c_n$ in practice, for reasons
Sec.~\ref{sec:why} sets out. Its recurrence is also reported to be numerically
unstable below the natural gate strength and safe only in the strong-coupling
regime, which in the convention of \eqref{eq:gate} excludes the entire range
$\lambda<1/3$~\cite{MiattoQuesada2020}; the experiments cited above operate
well below that value.

For a Gaussian input the Fock amplitudes have not been given in explicit closed
form. When they are needed numerically, the standard approach is to work in a
truncated Fock space, building $\hat x$ in dimension $T$, cubing it, and
exponentiating the resulting matrix~\cite{Killoran2019}. The procedure is not
innocuous. Restricting the space also restricts the unbounded generator, so
what is exponentiated is not the generator confined to the subspace. Errors of
this kind have been documented for related
constructions~\cite{NumErrors2022}, and the dependence on the cutoff can be
non-monotonic and even sensitive to its parity~\cite{FiniteDim2026}, which
undermines the usual convergence test. The implementation taken as reference
here states the difficulty in its own documentation, noting that the gate it
constructs is unitary but ``does not implement an exact cubic phase
gate''~\cite{Killoran2019}. Yet exact amplitudes are what the heralding
probabilities and resource measures above are built from, and both depend on
precisely the small amplitudes that truncation damages most.

The present work supplies those amplitudes for an arbitrary pure single-mode
Gaussian input. The route rests on a single observation. An amplitude is best
evaluated in the representation that diagonalizes the operator, and for
$V(\lambda)$ that is the position basis. The position-space resolution of the
identity leaves an integral of a Hermite polynomial against a Gaussian carrying
a cubic phase, and all such integrals descend from one master integral, which
completing the cube reduces to a single Airy function. The outcome,
Eq.~\eqref{eq:result}, is closed and cutoff-free, covers squeezing along any
quadrature axis together with arbitrary displacement and boost, and needs two
Airy evaluations for an entire Fock profile, the remaining index dependence
being algebraic.

Section~\ref{sec:setup} states the amplitude, the obstruction, and the
insertion that removes it, and Sec.~\ref{sec:closed} evaluates the resulting
integral in closed form. Section~\ref{sec:why} measures the cost of assembling
the same amplitude from the known gate matrix elements.
Section~\ref{sec:stab} analyses the cancellation that governs evaluation at
weak nonlinearity, where the earlier recurrence was reported to fail, and
Sec.~\ref{sec:recur} derives a recurrence in the Fock index for Gaussian input
together with the regime in which it is stable. Section~\ref{sec:bench}
benchmarks the truncated matrix exponential, whose error is found not to be
monotone in the cutoff. Appendix~\ref{app:contour} justifies the contour
displacement and the analytic continuation behind the closed form,
Appendix~\ref{app:low} lists the lowest amplitudes explicitly, and
Appendix~\ref{app:recur} derives the recurrence.

We work with $\hbar=1$, $\hat x\ket{x}=x\ket{x}$,
$\ket{n}=(\ad)^{n}\ket{0}/\sqrt{n!}$ and $\hat x=(a+\ad)/\sqrt2$. The Airy
function is the entire solution of $\Ai''(z)=z\Ai(z)$ that decays as
$z\to+\infty$; for real $z$ it is given by the improper oscillatory integral
$\Ai(z)=\frac{1}{2\pi}\int_{-\infty}^{\infty}e^{i(t^{3}/3+zt)}dt$, which is not
absolutely convergent, and elsewhere by analytic continuation~\cite{DLMF}.

\section{The amplitude, the obstruction, and the insertion}
\label{sec:setup}

\subsection{What is wanted}

Let $\ket{G}$ be the Gaussian state on which the gate acts. Every pure
single-mode Gaussian state is, up to an overall phase, a displaced squeezed
vacuum, and its position representation is~\cite{Weedbrook2012}
\begin{equation}
\psi_G(x)=\braket{x}{G}=\left(\frac{\Rea\kappa}{\pi}\right)^{1/4}
\exp\!\left[-\frac{\kappa}{2}(x-x_0)^{2}+ip_0 x\right],
\label{eq:gaussstate}
\end{equation}
with $x_0=\langle\hat x\rangle$, $p_0=\langle\hat p\rangle$ real and $\kappa$
complex. The density $|\psi_G(x)|^{2}$ falls off as
$e^{-\Rea\kappa\,(x-x_0)^{2}}$, so normalizability requires $\Rea\kappa>0$, and
carrying out the Gaussian integral then fixes the prefactor and its exponent
$1/4$. The parameter $\kappa$ is real when the squeezing is along $\hat x$ or
$\hat p$ and complex otherwise. The object of interest is the Fock amplitude
\begin{equation}
c_n=\bra{n}V(\lambda)\ket{G},
\label{eq:want}
\end{equation}
together with the gate matrix elements $\bra{n}V(\lambda)\ket{k}$, which we
recover as the special case $\kappa=1$, $x_0=p_0=0$ with a Fock ket in place of
$\ket{G}$.

\subsection{Why the Fock basis alone fails}

For Gaussian operations the Fock matrix elements are multivariate Hermite
polynomials, and this structure underlies the efficient simulation of Gaussian
circuits~\cite{Dodonov1994,Hamilton2017,FockGaussian}. The cubic generator
admits no such representation, and the obstruction is already visible in the
matrix.

In the Fock basis $\hat x=(a+\ad)/\sqrt2$ is tridiagonal with entries growing
as $\sqrt n$, so $\hat x^{3}$ is supported on $m-n\in\{\pm1,\pm3\}$, four
nonzero diagonals in all, with entries growing as $n^{3/2}$. It is unbounded,
and along the boundary it couples the retained subspace to the discarded one.
The finite-Fock implementation of Ref.~\cite{Killoran2019} truncates $\hat x$
to its $T\times T$ corner $\hat x_T$ and cubes it, and the result is not the
corner of the cube. The matrix elements of $\hat x^{3}$ are sums over two
intermediate Fock indices, and $(\hat x_T)^{3}$ omits the paths that pass above
the cutoff. Because $\hat x$ is tridiagonal, a three-step path leaving the
retained subspace must visit index $T$ from $T-1$ and return, and because
$\hat x^{3}$ is odd under parity it cannot return to $T-1$ in three steps; the
only surviving endpoints are $T-2$ and $T-1$. The two matrices therefore differ
in the single pair of entries $(T-2,T-1)$ and $(T-1,T-2)$, where the exact
entry is $\sqrt{(T-1)/2}\,(3T-3)/2$ and the omitted contribution is
$\sqrt{(T-1)/2}\,T/2$, a fraction $T/(3T-3)\to1/3$ of the whole.

The consequence is that $\hat x_T^{3}$ is not the restriction of $\hat x^{3}$
to the subspace, and $\exp[i\lambda(\hat x_T)^{3}]$ is accordingly the exact
exponential of a different, bounded operator rather than an approximation to
$V(\lambda)$ on that subspace. Its unitarity conceals the error, since the
generator has already been altered before the exponential is taken.

The failure is one of representation, not of computation. In the position basis
the gate is diagonal,
\begin{equation}
V(\lambda)\ket{x}=e^{i\lambda x^{3}}\ket{x},\qquad
\melem{x}{V(\lambda)}{y}=e^{i\lambda x^{3}}\delta(x-y),
\end{equation}
and there is nothing to truncate.

\subsection{The insertion}

The bra in \eqref{eq:want} is a Fock state, while the operator between it and
$\ket{G}$ is diagonal in position. The two representations are joined by the
spectral resolution of $\hat x$,
\begin{equation}
\int_{-\infty}^{\infty}\!dx\,\ketbra{x}{x}=\mathbb I ,
\label{eq:resolution}
\end{equation}
which, inserted into \eqref{eq:want} immediately to the left of the gate,
carries the amplitude into the basis adapted to the operator and gives
\begin{equation}
\bra{n}V(\lambda)\ket{G}
=\int_{-\infty}^{\infty}\!dx\;\braket{n}{x}\,e^{i\lambda x^{3}}\,\braket{x}{G}.
\label{eq:insert}
\end{equation}
The gate has become a scalar under the integral sign, and the operator content
of the amplitude now resides entirely in the two overlaps $\braket{n}{x}$ and
$\braket{x}{G}$, both of which are known in closed form.

Two remarks make the step precise. First, \eqref{eq:resolution} is the spectral
resolution of the self-adjoint operator $\hat x$ rather than an expansion in
Hilbert-space vectors, since $\ket{x}$ lies outside the space. The equality
holds in the sense of the projection-valued measure of $\hat x$, and this is
what licenses the insertion. Second, although $\hat x$ and $\hat x^{3}$ are
both unbounded, the domain questions that would ordinarily accompany them do
not arise here. The function $e^{i\lambda x^{3}}$ is a bounded Borel function
of a real variable, so the functional calculus defines $V(\lambda)$ as a
bounded unitary operator on the whole space, and \eqref{eq:insert} is exact as
written. Its integral converges absolutely, since the cubic phase has unit
modulus and Cauchy--Schwarz then bounds the integrand by the product of two
normalized wave functions,
\begin{equation}
\int\big|\braket{n}{x}e^{i\lambda x^{3}}\braket{x}{G}\big|\,dx
\le\big\|\braket{n}{\cdot}\big\|_{2}\,\big\|\braket{\cdot}{G}\big\|_{2}=1 .
\label{eq:cs}
\end{equation}

With $\braket{n}{x}=\mathcal N_n H_n(x)e^{-x^{2}/2}$,
$\mathcal N_n=(2^{n}n!)^{-1/2}\pi^{-1/4}$, and \eqref{eq:gaussstate},
Eq.~\eqref{eq:insert} becomes
\begin{equation}
\bra{n}V(\lambda)\ket{G}
=\mathcal N_n\mathcal N_G\int_{-\infty}^{\infty} H_n(x)\,
e^{-ax^{2}+bx+i\lambda x^{3}}\,dx,
\label{eq:fockME}
\end{equation}
with
\begin{equation}
a=\frac{1+\kappa}{2},\quad
b=\kappa x_0+ip_0,\quad
\mathcal N_G=\Big(\tfrac{\Rea\kappa}{\pi}\Big)^{\!1/4}\!e^{-\kappa x_0^{2}/2}.
\label{eq:ab}
\end{equation}
The input state has thus been compressed into two parameters. It enters the
integrand only through $a$ and $b$, and its remaining influence is the overall
constant $\mathcal N_G$. Since $\Rea\kappa>0$, the quadratic coefficient obeys
$\Rea a>\tfrac12$, which secures the Gaussian decay of the integrand for every
admissible input. The parameters $b$ and $\mathcal N_G$ are real only when the
squeezing lies along $\hat x$ or $\hat p$ and the state is undisplaced;
otherwise they are complex, and the closed form below is constructed to cover
that case.

Equation~\eqref{eq:fockME} holds without truncation, the Fock index entering
only through the degree of the polynomial under the integral sign. What
separates it from the gate matrix element treated in
Ref.~\cite{MiattoQuesada2020} is the departure from $a=1$ and $b=0$, that is,
the presence of squeezing, displacement and boost in the input.

Because $H_n$ is a polynomial of degree $n$, the integral in \eqref{eq:fockME}
is a finite linear combination of the moments
\begin{equation}
I_m(a,b,\lambda)=\int_{-\infty}^{\infty}\! x^{m}\,
e^{-ax^{2}+bx+i\lambda x^{3}}dx ,\;\; m\ge0,
\label{eq:moments}
\end{equation}
with coefficients read off from $H_n$, so evaluating the $I_m$ evaluates the
amplitude. The moments in turn follow from a single master integral,
\begin{equation}
G(a,b,\lambda)=\int_{-\infty}^{\infty}e^{-ax^{2}+bx+i\lambda x^{3}}\,dx,
\qquad
I_m=\frac{\partial^{m}G}{\partial b^{m}},
\label{eq:master}
\end{equation}
because $b$ appears linearly in the exponent, so each derivative in $b$ brings
down one factor of $x$. Passing the derivatives under the integral sign is
justified for every $m$ by dominated convergence. The cubic phase has unit
modulus, whence
$|x^{m}e^{-ax^{2}+bx+i\lambda x^{3}}|\le|x|^{m}e^{-\Rea a\,x^{2}+\Rea b\,x}$,
and on any bounded set of $b$ this majorant is integrable and independent of
$b$. The amplitude is therefore reduced to the evaluation of $G$, which is
carried out in Sec.~\ref{sec:closed}.

The same insertion gives the matrix elements of the gate. Taking the ket to be
a Fock state and using $\braket{x}{k}=\mathcal N_kH_k(x)e^{-x^{2}/2}$,
\begin{equation}
\bra{n}V(\lambda)\ket{k}
=\mathcal N_n\mathcal N_k\int_{-\infty}^{\infty} H_n(x)H_k(x)\,
e^{-x^{2}+i\lambda x^{3}}\,dx,
\label{eq:fockfock}
\end{equation}
which is \eqref{eq:fockME} with a product of two Hermite polynomials, of total
degree $n+k$, in place of one, obtained from the moments \eqref{eq:moments} by
setting $a=1$ and $b=0$ after the differentiation. This is Eq.~(130) of
Ref.~\cite{MiattoQuesada2020}, and its closed evaluation below reproduces their
Eq.~(148); we include it because the benchmark of Sec.~\ref{sec:bench} and the
comparison of Sec.~\ref{sec:why} both require it.

\section{Closed-form evaluation}
\label{sec:closed}

\subsection{The generating integral $G$ is an Airy function}

The generating integral \eqref{eq:master} is elementary once the quadratic term
is removed. Substituting $x=y+\sigma$ and choosing
\begin{equation}
3i\lambda\sigma-a=0
\quad\Longrightarrow\quad
\sigma=\frac{-ia}{3\lambda},
\label{eq:sigma}
\end{equation}
the exponent becomes
$i\lambda y^{3}+(b-a\sigma)y+(-\tfrac{2}{3}a\sigma^{2}+b\sigma)$, with no term
in $y^{2}$. What is left is the Airy integral,
\begin{equation}
\int e^{i\lambda y^{3}+\beta y}\,dy
=\frac{2\pi}{(3\lambda)^{1/3}}\,
\Ai\!\left(\frac{-i\beta}{(3\lambda)^{1/3}}\right),
\label{eq:airyint}
\end{equation}
and therefore
\begin{equation}
\boxed{\;G(a,b,\lambda)=A_{0}\,\Ai(z_{0})\;}
\label{eq:Gclosed}
\end{equation}
with $\sigma$ from \eqref{eq:sigma} and
\begin{equation}
A_{0}=e^{-\frac{2}{3}a\sigma^{2}+b\sigma}\,\frac{2\pi}{(3\lambda)^{1/3}},
\qquad
z_{0}=\frac{-i(b-a\sigma)}{(3\lambda)^{1/3}}.
\label{eq:A0z0}
\end{equation}

The shift \eqref{eq:sigma} is purely imaginary, so the substitution moves the
contour off the real axis and requires justification, as does the extension of
\eqref{eq:airyint} to complex $\beta$, where the defining oscillatory integral
no longer converges. Appendix~\ref{app:contour} supplies both. The contour is
displaced by a rectangle argument whose vertical sides vanish in the limit,
which establishes \eqref{eq:Gclosed} for $a$ real and positive and $b$ purely
imaginary, and analytic continuation in $b$ and then in $a$ carries it to the
full range. Equation~\eqref{eq:Gclosed} thus holds for $\lambda>0$,
$b\in\mathbb C$ and $\Rea a>0$, which by \eqref{eq:ab} covers every state of
the family. The one-parameter case $a=1$, $b=0$ recovers the argument
$z_0=(3\lambda)^{-4/3}$ of Ref.~\cite{MiattoQuesada2020}; what is added here is
that the same reduction survives complex $a$ and nonzero $b$, with $z_0$ moving
off the positive real axis into the complex plane.

Negative gate strengths are reached by reflecting the integration variable.
Substituting $x\to-x$ in \eqref{eq:master} gives
\begin{equation}
G(a,b,\lambda)=G\big(a,-b,|\lambda|\big),
\qquad \lambda<0,
\label{eq:negl}
\end{equation}
so that \eqref{eq:Gclosed}--\eqref{eq:A0z0} apply unchanged with $|\lambda|$ in
place of $\lambda$ and $-b$ in place of $b$. The moments then follow as
$I_m(a,b,\lambda)=(-1)^{m}I_m(a,-b,|\lambda|)$, and the slope introduced below
becomes $w_A=\partial_bz_0=+i|3\lambda|^{-1/3}$. All gate strengths below are
positive.

\subsection{From the master integral to the amplitude}

The moments $I_m$ of \eqref{eq:moments} were expressed in \eqref{eq:master} as
$b$-derivatives of $G$, and \eqref{eq:Gclosed} makes $G$ a product of two
factors, each of which depends on $b$ elementarily. One has
$\partial_bA_0=\sigma A_0$, while $z_0$ is affine in $b$ with slope
$w_A=-i(3\lambda)^{-1/3}$, so that
$\partial_b^{\,j}\Ai(z_0)=w_A^{\,j}\Ai^{(j)}(z_0)$. Differentiating the product
$m$ times by the Leibniz rule gives
\begin{equation}
I_m=\frac{\partial^{m}}{\partial b^{m}}\big[A_0\Ai(z_0)\big]
=A_0\sum_{j=0}^{m}\binom{m}{j}\,\sigma^{m-j}\,w_A^{\,j}\,\Ai^{(j)}(z_0).
\label{eq:leibniz}
\end{equation}
The Airy derivatives do not proliferate, because $\Ai''(z)=z\Ai(z)$ reduces
every one of them to the pair $\Ai,\Ai'$. Writing
$\Ai^{(j)}(z)=p_j(z)\Ai(z)+q_j(z)\Ai'(z)$, differentiation of this identity
gives the recursion
\begin{equation}
p_{j+1}=p_j'+z\,q_j,\qquad q_{j+1}=p_j+q_j',
\label{eq:pq}
\end{equation}
starting from $p_0=1$, $q_0=0$ (and hence $p_1=0$, $q_1=1$). The $p_j,q_j$ are
polynomials with integer coefficients, generated once and reused; this
elimination is the one used in Eqs.~(151)--(152) of
Ref.~\cite{MiattoQuesada2020}.

Expanding the Hermite polynomial as $H_n(x)=\sum_{m=0}^{n}h^{(n)}_m x^{m}$ and
inserting \eqref{eq:leibniz} into \eqref{eq:fockME},
\begin{equation}
\boxed{\;
\bra{n}V(\lambda)\ket{G}=\mathcal N_n\mathcal N_G
\sum_{m=0}^{n}h^{(n)}_m\,I_m(a,b,\lambda)\;},
\label{eq:result}
\end{equation}
a sum of $n+1$ terms in which no cutoff has been introduced at any stage.
Collecting the two independent special-function values,
\begin{equation}
\bra{n}V(\lambda)\ket{G}=P_n\,\Ai(z_0)+Q_n\,\Ai'(z_0),
\label{eq:PQ}
\end{equation}
with
\begin{equation}
\begin{pmatrix}P_n\\[2pt]Q_n\end{pmatrix}
=\mathcal N_n\mathcal N_GA_0\!\sum_{m=0}^{n}\!h^{(n)}_m\!
\sum_{j=0}^{m}\!\binom{m}{j}\sigma^{m-j}w_A^{\,j}\!
\begin{pmatrix}p_j(z_0)\\[2pt]q_j(z_0)\end{pmatrix}\!,
\label{eq:PQdef}
\end{equation}
so that $P_n$ and $Q_n$ involve only powers of $\sigma$, $w_A$ and $z_0$,
binomial coefficients and the $h^{(n)}_m$. The argument $z_0$ depends on the
state and the gate strength but not on $n$, so only two Airy-function
evaluations are required for the entire profile; the remaining $n$-dependence
is algebraic, residing in $P_n$ and $Q_n$. A single Hermite polynomial appears
here, where the gate matrix element carries a product of two. Expanding that
product would leave a double sum over the two Hermite expansions, which
\eqref{eq:result} replaces by a single sum.

The gate matrix elements \eqref{eq:fockfock} follow in the same way, with
$H_nH_k$ in place of $H_n$ and $a=1$, $b=0$, so that
\begin{equation}
\sigma=\frac{-i}{3\lambda},
\qquad
z_0=(3\lambda)^{-4/3},
\label{eq:z0ff}
\end{equation}
real and positive, where $\Ai$ decays monotonically.
Appendix~\ref{app:low} lists the lowest cases explicitly.

\subsection{Checks}

Equation~\eqref{eq:result} is exact, so no numerical test can confirm it. What
a test can do is verify that the derivation has been carried through without
error and measure how the closed form behaves in ordinary floating-point
arithmetic. Three checks are reported. The first is analytic. The amplitude for
a coherent input is obtained independently, without Hermite algebra, and its
generating function is required to reproduce \eqref{eq:result} and
\eqref{eq:fockfock}. The second compares the special case \eqref{eq:fockfock}
with the published result for the gate matrix elements. The third evaluates
\eqref{eq:result} in double precision against high-precision quadrature of the
defining integral \eqref{eq:fockME}.

\paragraph{Coherent input.}
For $\ket{G}=\ket{\alpha}$ one has
$\braket{x}{\alpha}=\pi^{-1/4}\exp(-x^{2}/2+\sqrt2\alpha x-\alpha^{2}/2
-|\alpha|^{2}/2)$, and \eqref{eq:insert} applied with a coherent bra as well
gives the matrix element without any Hermite algebra,
\begin{equation}
\bra{\beta}V(\lambda)\ket{\alpha}
=\frac{e^{-\frac{1}{2}(\alpha^{2}+\beta^{*2})
-\frac{1}{2}(|\alpha|^{2}+|\beta|^{2})}}{\sqrt\pi}\,
G\big(1,\sqrt{2}(\alpha+\beta^{*}),\lambda\big),
\label{eq:coh}
\end{equation}
a single Airy function whose argument is affine in $\alpha+\beta^{*}$, times a
Gaussian prefactor. Multiplying \eqref{eq:coh} by
$e^{(|\alpha|^{2}+|\beta|^{2})/2}$ leaves a function jointly analytic in
$\alpha$ and $\beta^{*}$ which, as in the general construction of
Ref.~\cite{MiattoQuesada2020}, generates the Fock matrix elements,
\begin{equation}
e^{\frac{|\alpha|^{2}+|\beta|^{2}}{2}}\bra{\beta}V(\lambda)\ket{\alpha}
=\sum_{n,k}\frac{\beta^{*n}\alpha^{k}}{\sqrt{n!\,k!}}\,\bra{n}V(\lambda)\ket{k}.
\label{eq:genfun}
\end{equation}
Reading off the coefficient of $\beta^{*n}$ at fixed $\alpha$ isolates
$\bra{n}V(\lambda)\ket{\alpha}$, which must agree with \eqref{eq:result}
evaluated at the Gaussian parameters of a coherent state, and reading off the
coefficient of $\beta^{*n}\alpha^{k}$ isolates $\bra{n}V(\lambda)\ket{k}$,
which must agree with \eqref{eq:fockfock}. Both identities were verified
numerically for $|\alpha|,|\beta|\le2$. Truncating the double series at $34$
terms per index and working at $60$ digits, the relative discrepancy between
the two sides ranges from $1.6\times10^{-13}$ to $7\times10^{-8}$ across the
amplitudes tested. The residual is the tail of the truncated series rather than
an error of the closed form, since at $n=34$ and $|\beta|=2$ the coefficient
$|\beta|^{n}/\sqrt{n!}$ is already of order $10^{-9}$, and it falls as the
truncation is raised. One bookkeeping point enters here. Relative to
\eqref{eq:gaussstate} with $\kappa=1$, $x_0=\sqrt2\,\Rea\alpha$ and
$p_0=\sqrt2\,\Ima\alpha$, the coherent state $\ket{\alpha}$ carries the
symmetric-ordering phase $e^{-ix_0p_0/2}$, which must be restored before the
two parametrizations are compared.

\paragraph{The gate matrix elements.}
The special case \eqref{eq:fockfock} agrees with Eq.~(148) of
Ref.~\cite{MiattoQuesada2020}. Evaluating both expressions at $60$ working
digits and forming the difference at $200$, the relative discrepancy is at most
$8.7\times10^{-51}$ over $\lambda\in\{0.10,0.2357,0.4714\}$ and all index pairs
with $n+k\le13$, and it falls by roughly one decade per additional working
digit. The seeds $\bra{0}V\ket{0}$ and $\bra{1}V\ket{1}$ agree with
Eqs.~(138)--(139) of the same reference to $4.0\times10^{-60}$ over the same
gate strengths.

\paragraph{Floating-point behaviour.}
Table~\ref{tab:verify} compares two independent evaluations of the same
amplitude. The first is \eqref{eq:result}, with the moments from
\eqref{eq:leibniz}, the Airy derivatives from \eqref{eq:pq}, and $\sigma$,
$A_0$, $z_0$, $a$, $b$, $\mathcal N_G$ from \eqref{eq:sigma}, \eqref{eq:A0z0}
and \eqref{eq:ab}, all in double precision. The second is the defining integral
\eqref{eq:fockME}, evaluated by quadrature at $110$ digits and taken as the
reference. The input is specified by the Gaussian parameters $(\kappa,x_0,p_0)$
and the gate by $\lambda$, and no other input enters.

At $\lambda=0.47$ the two agree to between $10^{-16}$ and $10^{-12}$, and the
agreement is insensitive to the choice of $(\kappa,x_0,p_0)$. The first row
sits at the level of double-precision round-off, so entries of that size should
be read as round-off rather than as measured errors. At $\lambda=0.10$ the
agreement deteriorates by four orders at $\kappa=1$ and by seven at $\kappa=4$.
The formula has not changed; what fails is its evaluation, and
Sec.~\ref{sec:stab} identifies the mechanism and removes it.

\begin{table}[!htbp]
\caption{Relative error of the closed form \eqref{eq:result}, evaluated in
double precision, against quadrature of \eqref{eq:fockME} carried at $110$
digits. The Gaussian input \eqref{eq:gaussstate} is specified by
$(\kappa,x_0,p_0)$. Rows four to six lie outside the reach of the gate matrix
element alone, being respectively antisqueezed, displaced and boosted, and
quadratically phased.}
\label{tab:verify}
\begin{ruledtabular}
\begin{tabular}{ccccc}
$\lambda$ & $\kappa$ & $(x_0,p_0)$ & $n$ & relative error \\
\colrule
$0.47$ & $1$      & $(0,0)$       & $0$ & $3\times10^{-16}$ \\
$0.47$ & $1$      & $(0,0)$       & $3$ & $6.7\times10^{-14}$ \\
$0.47$ & $1$      & $(0,0)$       & $7$ & $9.2\times10^{-14}$ \\
$0.47$ & $1/4$    & $(0,0)$       & $7$ & $6.8\times10^{-15}$ \\
$0.47$ & $1$      & $(0.8,0.5)$   & $7$ & $1.1\times10^{-12}$ \\
$0.47$ & $1+i/2$  & $(0,0)$       & $7$ & $3.5\times10^{-14}$ \\
$0.10$ & $1$      & $(0,0)$       & $7$ & $1.7\times10^{-9}$ \\
$0.10$ & $4$      & $(0,0)$       & $7$ & $2.2\times10^{-6}$ \\
\end{tabular}
\end{ruledtabular}
\end{table}

\section{Cost of the composition route}
\label{sec:why}

The gate matrix elements are available in closed
form~\cite{MiattoQuesada2020}, so the amplitude \eqref{eq:want} can in
principle be assembled from them by inserting a Fock resolution of the
identity,
\begin{equation}
\bra{n}V(\lambda)\ket{G}=\sum_{k=0}^{\infty}
\bra{n}V(\lambda)\ket{k}\braket{k}{G},
\label{eq:closure}
\end{equation}
and truncating the sum. The series converges, so the question is quantitative,
namely how many terms a given accuracy costs and how that cost behaves in the
parameter that matters. Let $K$ denote the highest Fock index retained, so that
truncating \eqref{eq:closure} at $K$ keeps the terms $k=0,\dots,K$; since
$\braket{k}{G}=0$ for odd $k$ on a squeezed vacuum, this amounts to $K/2+1$
nonvanishing terms. Two effects govern the cost, one in the number of terms and
one in the precision each term must carry.

\begin{figure}[!tb]
\includegraphics[width=\columnwidth]{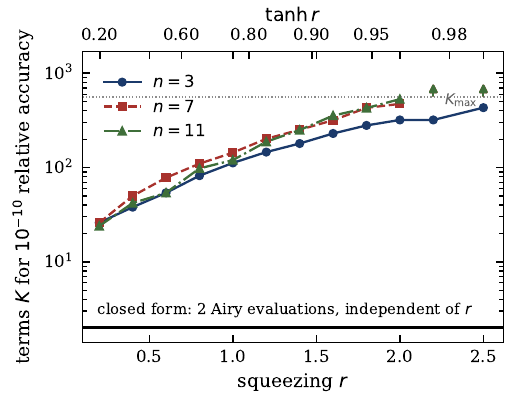}
\caption{Cost of the composition route \eqref{eq:closure}. Plotted is the
highest retained Fock index $K$ needed to reach a relative accuracy of
$10^{-10}$ in the amplitude $c_n$ of the finite-energy cubic phase state
\eqref{eq:cps0} at $\gamma=0.5$ ($\lambda=0.2357$), against the squeezing $r$
of the input, for three Fock indices; the upper axis gives the corresponding
$\tanh r$, which sets the convergence rate per retained pair of Fock indices.
The sum was carried at $140$ working digits so that the reported $K$ measures
the convergence of the series and not the arithmetic. Arrows mark points that
had not converged at $K_{\max}=560$. The heavy horizontal line is the cost of
the closed form \eqref{eq:result}, which is two Airy evaluations for the entire
profile and does not depend on $r$, on $n$, or on the target accuracy.}
\label{fig:closure}
\end{figure}

For the resource state \eqref{eq:cps0} the overlaps $\braket{k}{G}$ are those
of a squeezed vacuum. They vanish for odd $k$ and carry the factor
$(\tanh r)^{m}$ at $k=2m$, together with an algebraic prefactor that behaves as
$k^{-1/4}$, so that in the Fock index the decay is geometric at the rate
$(\tanh r)^{k/2}$. The matrix elements $\bra{n}V\ket{k}$, by contrast, do not
decay in $k$ over any comparable range, unitarity fixing only
$\sum_k|\bra{n}V\ket{k}|^{2}=1$. The convergence of \eqref{eq:closure} is
therefore governed by $\tanh r$ per retained pair of Fock indices, and that
ratio approaches unity in exactly the limit in which the cubic phase state
approaches its ideal form.
Figure~\ref{fig:closure} makes this concrete. The cost rises by more than an
order of magnitude across the range and shows no sign of levelling. At the
modest squeezing $r=0.8$ used in Sec.~\ref{sec:bench}, a single amplitude at
$n=7$ already requires the gate matrix elements up to $K=110$, each of which is
a double Hermite sum; at $r=2$ the figure is $K=474$, and by $r=2.2$ the series
has not reached the target by $K=560$. The behaviour is not peculiar to one
Fock index, since the three curves track one another, so the difficulty belongs
to the input state rather than to the amplitude being asked for.

The second effect is a growing demand on working precision. The terms of
\eqref{eq:closure} alternate and the gate matrix elements cancel among
themselves at large $k$, so the precision needed to resolve the sum grows with
$K$. At $r=0.8$ and $K=110$, the sum reaches its $10^{-10}$ target at $30$
working digits, is wrong by $1.5\times10^{-2}$ at $20$ digits, and at the $16$
digits of double precision overshoots the amplitude by a factor of
$1.6\times10^{3}$. The requirement follows the same degree-based estimate
established in Sec.~\ref{sec:stab}, applied with the total polynomial degree
$n+K$ and with $\sigma$ evaluated at $a=1$, as the gate matrix elements
require. At $K=560$ that estimate calls for some $85$ digits to carry the terms
alone, before the target accuracy is added, and correspondingly a sweep carried
at $25$ digits fails to converge at any $r$ in Fig.~\ref{fig:closure}.

The composition route thus reinstates, in a different guise, the cutoff that
the position-basis insertion was introduced to avoid, since one is again
obliged to choose a Fock truncation and to justify it, and it adds a precision
requirement of its own. Two structural differences complete the comparison. The
route is available only when the input has been expanded in Fock states, which
for complex $\kappa$, that is for squeezing along a general axis, requires the
two-index Gaussian coefficients rather than the elementary squeezed-vacuum
series. And it delivers one amplitude at a time, whereas \eqref{eq:PQ} shares
the argument $z_0$ across the whole profile.

\section{Numerical stability at weak nonlinearity}
\label{sec:stab}

The last two rows of Table~\ref{tab:verify} show the evaluation losing accuracy
as $\lambda$ decreases, and that is the regime the experiments occupy. The
demonstrated gate corresponds to $\lambda\approx0.17$~\cite{Sakaguchi2023} and
the decomposition schemes use strengths down to
$\lambda\simeq0.035$~\cite{Budinger2024}, a range sampled by the entries at
$\lambda=0.10$ and $\lambda=0.05$ in Table~\ref{tab:stab}. It is also the
regime that Ref.~\cite{MiattoQuesada2020} marks as unsafe for its recurrence,
so how to evaluate a cubic-phase Fock amplitude at experimentally relevant
strength has remained without a settled answer. The loss of accuracy is a
cancellation, it has an identifiable source, and it is removed at negligible
cost.

Two cancellations act together. The first is visible in \eqref{eq:sigma},
where the shift $\sigma=-ia/3\lambda$ diverges as $\lambda\to0$, so that the
Leibniz sum \eqref{eq:leibniz} contains terms of magnitude $|\sigma|^{m}$ while
the sum is $O(1)$. The second is the alternating sign pattern of the Hermite
coefficients $h^{(n)}_m$ in \eqref{eq:result}, which subtracts contributions of
comparable size. Take the third row of Table~\ref{tab:stab}, at $\lambda=0.10$
and $\kappa=4$. There $a=(1+\kappa)/2=5/2$ by \eqref{eq:ab}, $|\sigma|=8.33$,
and at $m=9$ the individual terms of \eqref{eq:leibniz} reach $3\times10^{8}$.
The first mechanism alone would leave eight of the sixteen digits of double
precision; the two together leave about four, which is the $1.4\times10^{-4}$
reported there.

\begin{table}[!htbp]
\caption{Relative error of \eqref{eq:result}, evaluated first in double
precision and then at $60$ digits, against quadrature of \eqref{eq:fockME}
carried at $110$ digits. The state \eqref{eq:gaussstate} has $x_0=p_0=0$.}
\label{tab:stab}
\begin{ruledtabular}
\begin{tabular}{ccccc}
$\lambda$ & $\kappa$ & $n$ & double & $60$ digits \\
\colrule
$0.10$ & $4$ & $5$ & $6.7\times10^{-9}$ & $8.1\times10^{-54}$\\
$0.10$ & $4$ & $7$ & $2.2\times10^{-6}$ & $2.5\times10^{-51}$\\
$0.10$ & $4$ & $9$ & $1.4\times10^{-4}$ & $3.8\times10^{-50}$\\
$0.05$ & $4$ & $7$ & $8.7\times10^{-4}$ & $7.2\times10^{-50}$\\
$0.05$ & $4$ & $9$ & $2.9\times10^{-3}$ & $5.5\times10^{-47}$\\
$0.05$ & $1$ & $9$ & $5.4\times10^{-4}$ & $2.0\times10^{-50}$\\
\end{tabular}
\end{ruledtabular}
\end{table}

The remedy is to carry \eqref{eq:leibniz} and \eqref{eq:pq} in extended
precision~\cite{mpmath}, and nothing else changes. Table~\ref{tab:stab}
evaluates the same closed form twice, once in double precision and once at $60$
digits, against the same $110$-digit quadrature of \eqref{eq:fockME} used in
Table~\ref{tab:verify}. Sixty digits restore the error to the $10^{-50}$ level
across the whole range, so the difficulty is one of arithmetic rather than of
the formula. The evaluation then costs $2.2\,\mathrm{ms}$ per amplitude on a
single core of an Intel Xeon at $2.2\,\mathrm{GHz}$, using mpmath~1.3.0 under
CPython~3.11 without the optional \texttt{gmpy} backend. The figure is quoted
to show that the cost is small in absolute terms, not as a speed comparison
with the truncated matrix exponential, which produces a whole $T\times T$
matrix at once. In the opposite direction the problem disappears without
intervention, since as $\lambda$ grows, $\sigma$ shrinks, the sum stops
cancelling, and double precision is adequate, as the upper rows of
Table~\ref{tab:verify} show.

How much precision to request can be estimated from the term magnitudes in
\eqref{eq:leibniz}, which suggest that the working precision should exceed the
target accuracy by $n\log_{10}|\sigma|$ digits, that is, linearly in the Fock
index and logarithmically in $1/\lambda$. This is an empirical guide rather
than a bound. Starting from $60$ digits at $\kappa=4$ it estimates $51.7$,
$21.6$ and $12.0$ correct digits for $(\lambda,n)=(0.10,9)$, $(0.01,20)$ and
$(0.01,25)$, against $49.4$, $21.1$ and $10.3$ measured against the same
reference. Sixty digits are therefore comfortable at $n=20$ and
$\lambda=10^{-2}$ and begin to fail near $n=25$.

A failure of a different kind occurs further down in $\lambda$, and it is worth
separating from the cancellation above because it is an overflow rather than a
loss of significance. The prefactor $A_0$ of \eqref{eq:A0z0} carries
$e^{-\frac{2}{3}a\sigma^{2}}$, which at $a=1$ is $e^{2/27\lambda^{2}}$, while
$\Ai(z_0)$ carries the compensating decay $e^{-\frac{2}{3}z_0^{3/2}}$ with
$\frac{2}{3}z_0^{3/2}=2/27\lambda^{2}$. The two cancel exactly, but each
separately leaves the range of double precision at $\lambda\simeq0.0102$, below
which the product silently returns zero and then \texttt{NaN} rather than
degrading gradually. This threshold is a factor of three below the smallest
strength in Ref.~\cite{Budinger2024} and so does not affect the results
reported here, but any implementation intended for weaker gates should form the
product $A_0\Ai(z_0)$ through the scaled Airy function
$\Ai(z)e^{\frac{2}{3}z^{3/2}}$, which removes the overflow at no cost.

\section{A recurrence in the Fock index}
\label{sec:recur}

The alternating sum of \eqref{eq:result} can be bypassed altogether. Writing
$J_n=\int H_n(x)e^{-ax^{2}+bx+i\lambda x^{3}}dx$, so that
$c_n=\mathcal N_n\mathcal N_G J_n$, the vanishing of
$\int\frac{d}{dx}\big[H_ne^{\phi}\big]dx$ with
$\phi'=3i\lambda x^{2}-2ax+b$ yields
\begin{align}
\frac{3i\lambda}{4}J_{n+2}=\;&aJ_{n+1}
-\Big(b+\tfrac{3}{2}i\lambda(2n+1)\Big)J_n \nonumber\\
&+2n(a-1)J_{n-1}-3i\lambda\,n(n-1)J_{n-2},
\label{eq:Jrec}
\end{align}
with $J_{-1}=J_{-2}=0$, so that $J_0$ and $J_1$ from \eqref{eq:leibniz} seed
the whole profile. The derivation is given in Appendix~\ref{app:recur}. At
$a=1$, $b=0$ the relation reduces to the recurrence of
Ref.~\cite{MiattoQuesada2020}, of which \eqref{eq:Jrec} is the extension to
arbitrary Gaussian input.

Where \eqref{eq:Jrec} may be used is a numerical question, and its stability
inherits the same $1/\lambda$ amplification analysed in Sec.~\ref{sec:stab}.
Seeded from \eqref{eq:leibniz} in double precision and run forward to $n=13$,
it reproduces \eqref{eq:result} to $1.1\times10^{-14}$ at $\lambda=0.4714$,
$\kappa=1$ and to $6.4\times10^{-13}$ at $\lambda=0.2357$, $\kappa=0.202$, that
is for the resource state of Sec.~\ref{sec:bench}, but to only
$6.3\times10^{-4}$ at $\lambda=0.10$, $\kappa=4$, and it diverges outright at
$\lambda=0.01$. Whether a backward, Miller-type application of
\eqref{eq:Jrec}, or an asymptotic expansion of $\Ai$ adapted to large $|z_0|$,
would recover that corner we have not determined, and the instability reported
in Ref.~\cite{MiattoQuesada2020} survives the extension.

The practical consequence is a division of labour, subject to the caveat that
we have surveyed parameters rather than proved bounds, so that an
implementation should verify the regime it is used in rather than rely on the
division. The extended-precision machinery of Sec.~\ref{sec:stab} is what the
strongly antisqueezed inputs and the sub-experimental gate strengths of
Table~\ref{tab:stab} require. At the resource-state parameters, however, the
recurrence suffices on its own. For $\lambda=0.2357$, $\kappa=0.202$ and
$n\le13$, that is \eqref{eq:cps0} at $r=0.8$ and the demonstrated gate
strength, \eqref{eq:Jrec} in ordinary double precision reproduces the closed
form to twelve digits, and a whole Fock profile follows from two seed values
and a linear pass.

\section{Benchmark against truncated evaluation}
\label{sec:bench}

The closed form is now compared against the finite-Fock construction of
Ref.~\cite{Killoran2019}, which builds $\hat x$ in a Fock space of dimension
$T$, cubes it, and takes the matrix exponential. That implementation is taken
as representative because it is in wide use, and the behaviour reported below
follows from the construction rather than from any particular code. Gate
strengths are quoted in the convention of Ref.~\cite{Killoran2019}, where the
gate is written $\exp(i\gamma\hat x_{\rm SF}^{3}/3\hbar)$ with $\hbar=2$. In
that convention the vacuum satisfies
$\langle\hat x_{\rm SF}^{2}\rangle=\hbar/2=1$, whereas ours gives
$\langle\hat x^{2}\rangle=1/2$, so $\hat x_{\rm SF}=\sqrt2\,\hat x$ and
$\exp(i\gamma\hat x_{\rm SF}^{3}/6)=\exp(i\gamma\,2\sqrt2\,\hat x^{3}/6)$. The
two parameters are therefore related by
\begin{equation}
\lambda=\frac{\sqrt2}{3}\,\gamma\simeq0.4714\,\gamma ,
\label{eq:convert}
\end{equation}
so the demonstrated strength $\lambda\approx0.17$ is $\gamma\simeq0.36$, while
$\gamma=0.5$ and $\gamma=1$ correspond to $\lambda=0.2357$ and
$\lambda=0.4714$.

\subsection{The gate}

\begin{figure}[!tb]
\includegraphics[width=\columnwidth]{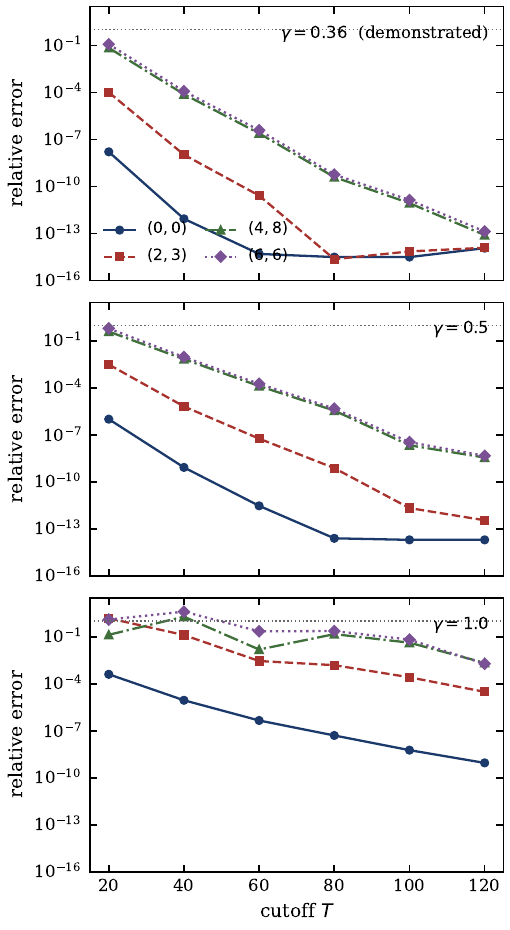}
\caption{Relative error of the truncated gate matrix element
$\bra{n}V\ket{k}$, obtained by cubing the $T\times T$ corner of $\hat x$ and
exponentiating, against the closed form \eqref{eq:fockfock}, as a function of
the cutoff $T$. From top to bottom, $\gamma=0.36$ ($\lambda=0.170$, the
demonstrated strength~\cite{Sakaguchi2023}), $\gamma=0.5$ ($\lambda=0.2357$)
and $\gamma=1.0$ ($\lambda=0.4714$), the two conventions being related by
\eqref{eq:convert}. Legend entries are the index pairs $(n,k)$, and the dotted
line marks $100\%$ error. In the upper two panels the error falls monotonically
at every index pair, and in the top panel it reaches the double-precision floor
by $T=80$. In the lower panel the pairs $(4,8)$ and $(6,6)$ do not, the error
rising between $T=20$ and $T=40$ and again between $T=60$ and $T=80$, so that
increasing the cutoff does not guarantee improved accuracy.}
\label{fig:gate}
\end{figure}

Figure~\ref{fig:gate} shows the relative error of the truncated matrix elements
$\bra{n}V\ket{k}$ against \eqref{eq:fockfock}, for four index pairs, six
cutoffs and three gate strengths spanning the demonstrated value. Three
features stand out.

Errors grow rapidly with the Fock indices at fixed cutoff, so the accuracy of
the low-lying corner of the matrix says nothing about the rest. They grow with
the gate strength, so the useful cutoff depends on the circuit and not on the
mode alone. And at the two larger strengths they are not monotone in $T$. At
$\gamma=1$ the element $\bra{4}V\ket{8}$ carries a $13\%$ error at $T=20$, is
wrong by $192\%$ at $T=40$, falls to $1.5\%$ at $T=60$, rises again to $15\%$
at $T=80$, and reaches $2\times10^{-3}$ only at $T=120$; the diagonal element
$\bra{6}V\ket{6}$ oscillates likewise, from $126\%$ to $402\%$ to $22\%$ across
the first three cutoffs. At $\gamma=0.5$ the same two pairs are the slowest to
converge, though still monotonically.

Nothing forbids this behaviour. The truncated exponentials are exponentials of
different operators for different $T$, and no variational principle orders
them. It does mean that monotonic improvement with the cutoff cannot be
assumed, so a finite-cutoff convergence study can give misleading evidence of
accuracy. This is the practical form, for the cubic gate, of the parity and
self-adjointness effects reported in Ref.~\cite{FiniteDim2026}.

The three strengths were chosen to span the demonstrated value rather than to
sit at it, and the effects separate there. At $\gamma\simeq0.36$ the error
falls monotonically at every index pair examined and reaches the
double-precision floor by $T=80$, so the non-monotonic behaviour is a strong-coupling phenomenon and is not
claimed at the demonstrated strength. What survives there is the size of the
error at small cutoff, $12\%$ at $T=20$ for $\bra{6}V\ket{6}$, and this is a
per-gate figure in schemes that apply the gate in long sequences, where such
errors accumulate. For the index pairs studied here convergence is eventually
observed, but only at cutoffs where the $O(T^{3})$ cost of the dense matrix
exponential that builds each gate becomes the dominant expense of the
simulation, and we make no claim about the cutoff required in general.

\subsection{The finite-energy cubic phase state}

\begin{figure}[!tb]
\includegraphics[width=\columnwidth]{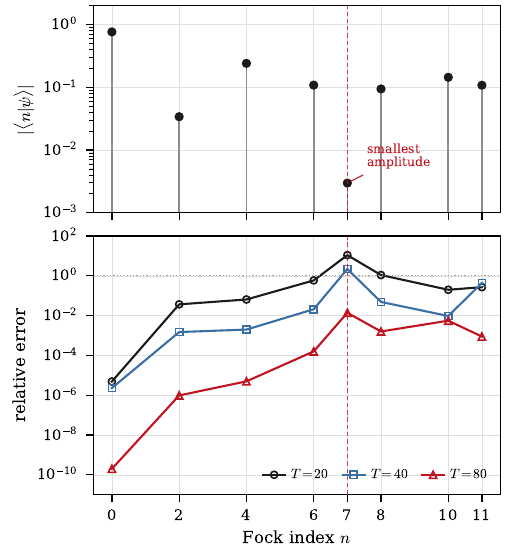}
\caption{The finite-energy cubic phase state \eqref{eq:cps0} at $r=0.8$, that
is $\kappa=e^{-2r}=0.202$, and $\gamma=0.5$ ($\lambda=0.2357$). The upper panel
gives the exact Fock amplitudes from \eqref{eq:result}. The lower panel gives
the relative error of the truncated evaluation, obtained by cubing the
$T\times T$ corner of $\hat x$ and exponentiating, at cutoffs $T=20,40,80$,
with the dotted line marking $100\%$ error. The dashed vertical guide marks
$n=7$, where the upper panel shows the smallest exact amplitude and the lower
panel the largest error at every cutoff.}
\label{fig:cps}
\end{figure}

The state of practical interest is \eqref{eq:cps0}, with $S(r)$ in the
convention that stretches the position quadrature,
$S\dg(r)\hat xS(r)=e^{r}\hat x$. Its wave function is then
\eqref{eq:gaussstate} with $\kappa=e^{-2r}$ real and $x_0=p_0=0$, so
\eqref{eq:result} applies with $a=(1+e^{-2r})/2$ and $b=0$. The identification
is fixed by the second moment, since \eqref{eq:gaussstate} with $x_0=p_0=0$
gives $\langle\hat x^{2}\rangle=1/2\Rea\kappa$, while the stated convention
gives $\langle0|S\dg\hat x^{2}S|0\rangle=e^{2r}/2$, and the two agree precisely
when $\kappa=e^{-2r}$. Squeezing along a general axis makes $\kappa$ complex,
its imaginary part being the quadratic phase, and \eqref{eq:result} covers that
case unchanged. Exact amplitudes at $r=0.8$ and $\gamma=0.5$ are listed in
Table~\ref{tab:cps} and plotted in the upper panel of Fig.~\ref{fig:cps}, whose
lower panel gives the relative error of the truncated evaluation.

\begin{table}[!tb]
\caption{Exact Fock amplitudes of the finite-energy cubic phase state
\eqref{eq:cps0} at $r=0.8$ and $\gamma=0.5$ ($\lambda=0.2357$), from
\eqref{eq:result} at $60$ digits, for use as reference values. All digits shown
are stable under an independent quadrature carried at $110$ digits. The cubic
phase destroys the parity of the squeezed vacuum, so the odd amplitudes do not
vanish and are listed. As a check on the set,
$\sum_{n\le11}|\braket{n}{\psi}|^{2}=0.8888693$.}
\label{tab:cps}
\begin{ruledtabular}
\begin{tabular}{cccc}
$n$ & $|\braket{n}{\psi}|$ & $n$ & $|\braket{n}{\psi}|$ \\
\colrule
$0$ & $0.7677890$ & $6$  & $0.1087204$ \\
$1$ & $0.2993510$ & $7$  & $0.0029840$ \\
$2$ & $0.0340905$ & $8$  & $0.0944669$ \\
$3$ & $0.1810120$ & $9$  & $0.1429172$ \\
$4$ & $0.2417572$ & $10$ & $0.1449824$ \\
$5$ & $0.2084738$ & $11$ & $0.1082812$ \\
\end{tabular}
\end{ruledtabular}
\end{table}

The two panels should be read together. The truncation error tracks the
smallness of the amplitude. At $n=7$, where the exact amplitude dips to
$3\times10^{-3}$, the truncated value is wrong by a factor of eleven at $T=20$
and still by a factor of two at $T=40$, while its neighbours at the same cutoff
are accurate to a few percent. Small amplitudes are precisely those on which
heralding probabilities and fidelity measures depend, so the truncated
calculation is least accurate where the applications are most sensitive. Nor is
the error uniformly smaller at $T=40$ than at $T=20$, as the behaviour at
$n=11$ shows.

\section{Discussion}
\label{sec:discussion}

The Airy structure of cubic-phase problems is
old~\cite{GhoseSanders2007,Moore2025}, and its Fock-basis form for the gate
matrix elements is due to Ref.~\cite{MiattoQuesada2020}. Four things are added
here.

The first is Eq.~\eqref{eq:result}, a closed-form expression for the Fock
amplitudes of the cubic phase gate acting on an arbitrary pure single-mode
Gaussian state, with squeezing along an arbitrary quadrature axis and arbitrary
displacement and momentum boost, together with the evidence of
Sec.~\ref{sec:why} that this is not obtained in practice by composing the known
matrix elements. The second is a stable extended-precision evaluation of that
expression in the regime of weak nonlinearity where the recurrence of
Ref.~\cite{MiattoQuesada2020} was reported to become unstable. The third is the
extension \eqref{eq:Jrec} of that recurrence to Gaussian input, with a
numerical characterization of its stable operating regime. The fourth is the
benchmark of Sec.~\ref{sec:bench}, whose main lesson is that above the
demonstrated gate strength the error of the truncated construction is not
monotone in the cutoff for several of the matrix elements examined, so that
monotonic improvement with the cutoff cannot be assumed in general. At the
demonstrated strength the error is monotone, but reaches $12\%$ at the smallest
cutoff tested.

Three limitations should be stated plainly.

\emph{A single gate.} The formula covers one cubic gate acting on a Gaussian
state. Two gates in succession, separated by a rotation, lead to a double
Airy-type integral that we have not reduced in closed form.

\emph{Precision.} Extended precision is needed for the strongly antisqueezed
inputs and the very weak gate strengths of Table~\ref{tab:stab}. At the
resource-state parameters tested in Sec.~\ref{sec:recur} double precision
through \eqref{eq:Jrec} was sufficient. We have not mapped the boundary between
the two regimes, and the remaining corner is a target for a backward recurrence
or an adapted asymptotic expansion.

\emph{Multimode gates.} The treatment is single-mode, but the insertion of
Sec.~\ref{sec:setup} applies verbatim to the multimode gates that carry the
optical decomposition schemes, namely the cubic quantum-nondemolition gate
$e^{i\alpha\hat x_1\hat x_2^{2}}$ and the continuous-variable Toffoli gate
$e^{i\beta\hat x_1\hat x_2\hat x_3}$~\cite{Budinger2024,Hanamura2024}, these
being functions of commuting quadratures as well. For the
quantum-nondemolition gate the reduction goes one step further than in the
general case, because $\hat x_1$ enters the exponent quadratically and
integrates out in closed form, leaving the one-dimensional integral
$\int dx_2\,e^{-Ax_2^{4}+Bx_2^{2}+Cx_2}$ with $A=\alpha^{2}/4a_1$, whose real
part is positive since $\Rea a_1>0$. Writing $F$ for that integral, integration
by parts gives $4A\,\partial_C^{3}F=2B\,\partial_CF+CF$ in place of
$\Ai''(z)=z\Ai(z)$, while $\partial_AF=-\partial_C^{4}F$ and
$\partial_BF=\partial_C^{2}F$ reduce the remaining parameter derivatives to
$\partial_C$ as well. The elimination of \eqref{eq:pq} then applies with a
third-order relation in place of a second-order one, so that every
$C$-derivative becomes a polynomial combination of $F$, $\partial_CF$ and
$\partial_C^{2}F$, and an amplitude collapses to three special-function values
of Pearcey type at one argument rather than two of Airy type. For $C=0$ the quartic integral
degenerates to the Hankel form obtained for the single-mode quartic phase gate
in Appendix~E of Ref.~\cite{MiattoQuesada2020}. The Toffoli gate also reduces
to one dimension, $\hat x_1$ and then $\hat x_2$ integrating out in turn, but
the surviving integrand carries an algebraic prefactor and a rational exponent,
for which no such closure exists. We record these reductions here and develop
them elsewhere.

The immediate applications are to the exact evaluation of success probabilities
in heralded cubic-phase-state preparation, where Fig.~\ref{fig:cps} shows that
the amplitudes most damaged by truncation are the small ones on which those
probabilities depend, and to the benchmarking of CV simulators, for which the
closed form supplies reference values at any Fock index without a convergence
study in the cutoff.

\appendix
\section{Contour displacement and analytic continuation}
\label{app:contour}

This appendix justifies the two analytic steps behind \eqref{eq:Gclosed}, the
imaginary shift \eqref{eq:sigma} and the use of \eqref{eq:airyint} at complex
argument.

\subsection{The shifted contour}

Take $a$ real and positive and $b$ purely imaginary, so that the shift
$\sigma=-ia/3\lambda$ is purely imaginary and carries the contour onto the line
$\Ima y=a/3\lambda$. Writing $y=u+iv$,
\begin{equation}
|e^{i\lambda y^{3}}|=e^{-3\lambda u^{2}v+\lambda v^{3}},
\end{equation}
which is Gaussian in $u$ for $v>0$ but degenerates at $v=0$. The integrand
therefore does not decay uniformly across the strip, and the displacement
cannot be justified by a uniform bound. It is enough to control the vertical
segments of the rectangle. At $\Rea y=\pm R$ the cubic and linear factors
together obey
\begin{equation}
|e^{i\lambda y^{3}+\beta y}|
=e^{-3\lambda R^{2}v+\lambda v^{3}}\,e^{-\Ima\beta\,v}
\le Ce^{-3\lambda R^{2}v},
\end{equation}
with $C=\sup_{0\le v\le a/3\lambda}e^{\lambda v^{3}-\Ima\beta\,v}$ finite and
independent of $R$, since $\beta$ is purely imaginary here and the segment has
fixed finite length. Integrating the bound over $v$ gives at most
$C/3\lambda R^{2}$, which vanishes as $R\to\infty$. The two contours therefore
agree, the real-axis integral existing as an improper oscillatory integral
because its integrand has modulus one there. In this regime $z_0$ is real, so
\eqref{eq:airyint} is used only where the Airy integral converges.

\subsection{Extension to complex parameters}

The general case follows by continuing in one variable at a time.

Fix $a>0$ real. Both sides of \eqref{eq:Gclosed} are entire in $b$, the left by
the domination noted after \eqref{eq:master}, which makes the integral
differentiable in $b$ to every order on any bounded set, and the right because
$\Ai$ is entire, $A_0$ is the exponential of an affine function of $b$, and
$z_0$ is affine in $b$. The two agree on the imaginary axis, a set with limit
points in $\mathbb C$, so the identity theorem gives agreement for every
$b\in\mathbb C$.

Now fix $b\in\mathbb C$. The right-hand side is analytic in $a$ on
$\Rea a>0$, since $\Ai$ is entire and $\sigma$, $A_0$, $z_0$ are analytic
there. So is the left. On any compact $K\subset\{\Rea a>0\}$,
\begin{equation}
|e^{-ax^{2}+bx+i\lambda x^{3}}|\le e^{-\alpha x^{2}+\Rea b\,x},
\qquad \alpha=\min_{a\in K}\Rea a>0,
\end{equation}
a bound independent of $a$ and integrable, so the integral is continuous on $K$
and its contour integral over any closed triangle in $K$ vanishes by Fubini,
and Morera's theorem then gives holomorphy. The two sides agree on the positive
real axis by the previous step, again a set with limit points in the
half-plane, and a second application of the identity theorem extends the
agreement to all $a$ with $\Rea a>0$, which is the range \eqref{eq:ab}
requires.

\section{The lowest amplitudes explicitly}
\label{app:low}

With $\sigma$, $w_A$, $A_0$ and $z_0$ as in Sec.~\ref{sec:closed}, the first
three moments \eqref{eq:leibniz} are
\begin{align}
I_0&=A_0\,\Ai(z_0),\nonumber\\
I_1&=A_0\left[\sigma\Ai(z_0)+w_A\Ai'(z_0)\right],\\
I_2&=A_0\left[(\sigma^{2}+w_A^{2}z_0)\Ai(z_0)+2\sigma w_A\Ai'(z_0)\right],
\nonumber
\end{align}
where $\Ai''(z_0)=z_0\Ai(z_0)$ has been used in $I_2$. Since $H_0=1$, $H_1=2x$
and $H_2=4x^{2}-2$, the first three amplitudes on the Gaussian
\eqref{eq:gaussstate} are
\begin{align}
\bra{0}V\ket{G}&=\mathcal N_0\mathcal N_G\,I_0,\nonumber\\
\bra{1}V\ket{G}&=\mathcal N_1\mathcal N_G\,2I_1,\\
\bra{2}V\ket{G}&=\mathcal N_2\mathcal N_G\,(4I_2-2I_0),\nonumber
\end{align}
with $a$, $b$ and $\mathcal N_G$ read off from the state through
\eqref{eq:ab}. The gate matrix elements \eqref{eq:fockfock} use the same
expressions evaluated at $a=1$ and $b=0$, where $\sigma=-i/3\lambda$,
$z_0=(3\lambda)^{-4/3}$ by \eqref{eq:z0ff} and
$A_0=2\pi(3\lambda)^{-1/3}e^{-2\sigma^{2}/3}$, with $H_nH_k$ in place of $H_n$.
For example $\bra{0}V\ket{1}=2\mathcal N_0\mathcal N_1I_1(1,0,\lambda)$, while
$\bra{0}V\ket{0}$ and $\bra{1}V\ket{1}$ are the diagonal cases checked against
Eqs.~(138)--(139) of Ref.~\cite{MiattoQuesada2020} in Sec.~\ref{sec:closed}.

\section{Derivation of the recurrence}
\label{app:recur}

Let $\phi(x)=-ax^{2}+bx+i\lambda x^{3}$, so that
$\phi'(x)=3i\lambda x^{2}-2ax+b$. The integrand of $J_n$ decays as
$e^{-\Rea a\,x^{2}}$ at both ends, so $H_n(x)e^{\phi(x)}$ vanishes there and
\begin{align}
0&=\int_{-\infty}^{\infty}\frac{d}{dx}\Big[H_n(x)e^{\phi(x)}\Big]dx
\nonumber\\
&=\int_{-\infty}^{\infty}\Big[H_n'(x)+H_n(x)\phi'(x)\Big]e^{\phi(x)}dx .
\label{eq:ibp}
\end{align}
The Hermite polynomials satisfy $H_n'=2nH_{n-1}$ together with
\begin{align}
xH_n&=\tfrac12H_{n+1}+nH_{n-1},
\label{eq:hmult1}\\
x^{2}H_n&=\tfrac14H_{n+2}+\big(n+\tfrac12\big)H_n+n(n-1)H_{n-2}.
\label{eq:hmult2}
\end{align}
Substituting \eqref{eq:hmult1} and \eqref{eq:hmult2} into \eqref{eq:ibp} and
collecting the coefficient of each $J_{n+j}$ gives
\begin{align}
\tfrac{3i\lambda}{4}J_{n+2}-aJ_{n+1}
&+\Big(b+\tfrac{3}{2}i\lambda(2n+1)\Big)J_n \nonumber\\
&-2n(a-1)J_{n-1}+3i\lambda n(n-1)J_{n-2}=0,
\label{eq:collected}
\end{align}
where the coefficient of $J_{n-1}$ combines the $2n$ from $H_n'$ with the
$-2an$ from the linear term of $\phi'$. Rearranging \eqref{eq:collected} gives
\eqref{eq:Jrec}.

\section*{Data availability}
The code that generates every table and figure in this paper is openly
available at Zenodo~\cite{ZenodoCode}. No other data were generated.


\end{document}